\documentclass{article}

\usepackage[utf8]{inputenc}
\usepackage[T1]{fontenc}

\PassOptionsToPackage{table, dvipsnames}{xcolor}
\usepackage{xcolor}

\PassOptionsToPackage{colorlinks=true, citecolor=blue, linkcolor=blue, urlcolor=black}{hyperref}

\usepackage{arxiv}
\usepackage[utf8]{inputenc}
\usepackage[T1]{fontenc}
\usepackage{url}
\usepackage{booktabs}
\usepackage{nicefrac}
\usepackage{microtype}
\usepackage{lipsum}
\usepackage{graphicx}
\usepackage{natbib}
\usepackage{doi}
\usepackage{algorithm}
\usepackage{multirow}
\usepackage{multicol}
\usepackage{amsmath,amssymb,amsfonts}
\usepackage{amsthm}
\usepackage{mathrsfs}
\usepackage{float}
\usepackage{algorithmic}
\usepackage{bm}
\usepackage{caption}
\usepackage{hyperref}

\title{DiffPINN: Generative diffusion-initialized physics-informed neural networks for accelerating seismic wavefield representation
}

\author{ \href{https://orcid.org/0000-0001-8868-7967}{\includegraphics[scale=0.06]{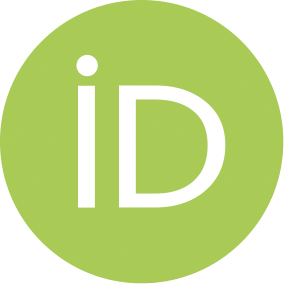}\hspace{1mm}Shijun~Cheng}\\
	Division of Physical Science and Engineering\\
	King Abdullah University of Science and Technology\\
	Thuwal 23955-6900, Saudi Arabia \\
	\texttt{sjcheng.academic@gmail.com} \\
        \And
	\href{https://orcid.org/0000-0002-9363-9799}{\includegraphics[scale=0.06]{orcid.png}\hspace{1mm}Tariq~Alkhalifah} \\
	Division of Physical Science and Engineering\\
	King Abdullah University of Science and Technology\\
	Thuwal 23955-6900, Saudi Arabia \\
	\texttt{tariq.alkhalifah@kaust.edu.sa} \\
}

\renewcommand{\shorttitle}{DiffPINN for accelerating seismic wavefield representation}

\hypersetup{
pdftitle={A template for the arxiv style},
pdfsubject={q-bio.NC, q-bio.QM},
pdfauthor={David S.~Hippocampus, Elias D.~Striatum},
pdfkeywords={First keyword, Second keyword, More},
}

\begin{document}
\maketitle

\begin{abstract}
Physics-informed neural networks (PINNs) offer a powerful framework for seismic wavefield modeling, yet they typically require time-consuming retraining when applied to different velocity models. Moreover, their training can suffer from slow convergence due to the complexity of of the wavefield solution. To address these challenges, we introduce a latent diffusion-based strategy for rapid and effective PINN initialization. First, we train multiple PINNs to represent frequency-domain scattered wavefields for various velocity models, then flatten each trained network’s parameters into a one-dimensional vector, creating a comprehensive parameter dataset. Next, we employ an autoencoder to learn latent representations of these parameter vectors, capturing essential patterns across diverse PINN's parameters. We then train a conditional diffusion model to store the distribution of these latent vectors, with the corresponding velocity models serving as conditions. Once trained, this diffusion model can generate latent vectors corresponding to new velocity models, which are subsequently decoded by the autoencoder into complete PINN parameters. Experimental results indicate that our method significantly accelerates training and maintains high accuracy across in-distribution and out-of-distribution velocity scenarios.
\end{abstract}

\section{\textbf{Introduction}}
 Seismic wavefield modeling is a crucial aspect of geophysical exploration, earthquake monitoring, and subsurface characterization \citep{carcione2002seismic}. Accurate modeling of wave propagation through complex subsurface structures enables better understanding of geological formations and improves seismic imaging and inversion results \citep{fichtner2010full}. Conventional numerical methods, such as finite-difference (FD) \citep{virieux1984sh, virieux1986p, moczo20023d, robertsson1994viscoelastic}, finite-element \citep{padovani1994low, koketsu2004finite}, and spectral-element methods \citep{zhu2014modeling, wang2022propagating}, have been widely used to simulate seismic wave propagation. While these methods often yield high-fidelity results, they typically demand significant computational resources, especially for large-scale, three-dimensional problems. Conventional numerical methods also suffer from discretization errors, especially if high-order derivatives are involved. Additionally, modeling wavefields under varying velocity conditions, like in inversion tasks, requires re-running these simulations, making the entire process time-consuming. High performance computing resources can mitigate some of these costs \citep{yang2015graphics, wang2019cu}, but the need for repeated and extensive simulations remains a significant bottleneck. 

Physics-informed neural networks (PINNs) \citep{raissi2019physics} have gained considerable attention as a powerful framework for solving partial differential equations (PDEs) in various fields. In the realm of seismic wavefield modeling, PINNs have shown great potential due to its grid-free and unsupervised features. \cite{alkhalifah2021wavefield} and \cite{song2021solving} pioneered the use of PINNs to solve the Helmholtz equation for representing frequency-domain scattered wavefields in isotropic and anisotropic media. \cite{bin2021pinneik} developed PINNs to approximate seismic traveltimes by embedding the Eikonal equation within the network’s loss function. \cite{song2021wavefield} proposed using PINNs for wavefield reconstruction inversion, where PINNs represent frequency wavefields to link observed data with velocity models within the domain of interest. \cite{rasht2022physics} used PINNs to solve the acoustic wave equation in the time domain to represent pressure wavefields, further demonstrating the potential of the trained PINN as a forward simulator for inversion. \cite{huang2022pinnup} proposed a frequency upscaling and neuron splitting strategy within a PINN framework to progressively simulate high-frequency scattered wavefields, effectively leveraging lower-frequency pretraining to significantly improve accuracy and convergence speed. To address challenges posed by nonsmooth media, \cite{wu2023helmholtz} introduced quadratic neuron activations and incorporated perfectly matched layer boundary conditions into a PINN framework, significantly enhancing the accuracy and convergence speed of frequency-domain acoustic and visco-acoustic wavefield simulations. \cite{alkhalifah2024physics} proposed integrating an adaptive Gabor-based hidden layer into PINNs, significantly improving computational efficiency and accuracy. \cite{chai2024modeling} proposed a PINN using multiscale Fourier feature mapping and adaptive activation functions to directly simulate multisource and multifrequency acoustic wavefields. \cite{cheng2024robust, cheng2025discovery} employed PINNs to reconstruct complete wavefields from sparse observations. Then, they leveraged the PINN framework, which provides physics-based criteria, to directly discover seismic wave equations from noisy and sparse observations. 

Essentially, PINNs learn to functionaly approximate the solution of the wave equation given specific physical constraints (e.g., initial/boundary conditions, frequencies, and velocity models) \citep{alkhalifah2021wavefield}. However, for seismic problems, the wavefield solution is highly sensitive to the velocity distribution within the subsurface. A change in the velocity model effectively alters the function that the PINN must approximate, since the spatiotemporal pattern of wave propagation depends on local variations in medium properties. This change in the underlying solution space typically necessitates retraining from scratch for each new velocity model, as a single set of PINN parameters tuned to one model cannot readily represent the distinct solution corresponding to another \citep{cheng2025meta}. Consequently, we can face significant computational overhead when repeatedly training PINNs for large-scale geophysical simulations. These challenges are further exacerbated by the slow convergence PINNs may exhibit for complex wavefields, underscoring the need for more efficient training approaches. 

To address the challenge of retraining PINNs for each new velocity model, some preliminary studies have already been developed to address the issue of PINN's adaptability in representing seismic wavefields for diverse velocity models. For example, \cite{taufik2024multiple} proposed a LatentPINN framework. They first trained an autoencoder on various velocity models using self-supervised reconstruction to obtain latent representation of the velocity models, which are then used as extra inputs to a PINN that learns to represent corresponding wavefields. Once trained, the PINN can directly predict wavefields for a new velocity model from a similar distribution without any further training. We \citep{cheng2025meta} proposed a novel meta-learning-based initialization for PINNs, where a common initial network is first trained using meta-learning across limited velocity models. The meta-trained initialization can rapidly adapt when applied to any new velocity model, significantly speeding up convergence and improving accuracy compared to vanilla PINNs with random initialization. Building upon this work, we further proposed a Meta-LRPINN framework \citep{cheng2025multi}, which integrates low-rank weight decomposition using singular value decomposition and a frequency embedding hypernetwork into the meta-learning approach. This new framework significantly accelerates convergence and improves accuracy for wavefield modeling across different frequencies and velocity models, while demonstrating strong computational efficiency. Our two studies demonstrated that a robust initialization of network parameters is crucial for improving the accuracy and convergence speed of PINNs, as it effectively prevents PINNs from spending excessive time in the early stages of optimization searching for a reliable direction due to random initialization. Therefore, this motivates us to develop a more powerful method to provide the initialization parameters of PINNs, so as to further improve their performances. 

Recently, \cite{wang2024neural} proposed Neural Network Diffusion, showing that an unconditional latent diffusion model can directly generate the final-layer parameters of vision networks, while matching or even exceeding SGD-trained models and accelerating fine-tuning by an order of magnitude. Inspired by this proof of concept, we extend the idea into the physics domain and propose a highly innovative concept—using a latent diffusion model to generate initialization parameters for PINNs. Specifically, we first train multiple PINNs on a range of velocity models and flatten their network parameters into one-dimensional vectors. An autoencoder then learns latent representations of these parameter vectors, capturing essential patterns shared across different wavefields. In parallel, a one-dimensional conditional diffusion model, conditioned on velocity models, is trained on these latent representations. When presented with a new velocity model, the diffusion model generates a latent vector that is subsequently decoded into a full set of PINN parameters. By starting from this physics-aware initialization, PINN training converges more rapidly while maintaining high accuracy. Experimental results demonstrate the effectiveness and efficiency of this method across diverse seismic wavefield scenarios.

\section{\textbf{Background}}
In this section, we present a concise review of the background knowledge involved in our approach. First, we discuss how PINNs are optimized to represent seismic wavefields. Next, we give a brief overview of generative diffusion models (GDMs), focusing on their forward and reverse processes. Finally, we highlight the conceptual connections between PINN optimization and the iterative denoising in GDMs. 

\begin{figure}[htbp]
\centering
\includegraphics[width=0.6\textwidth]{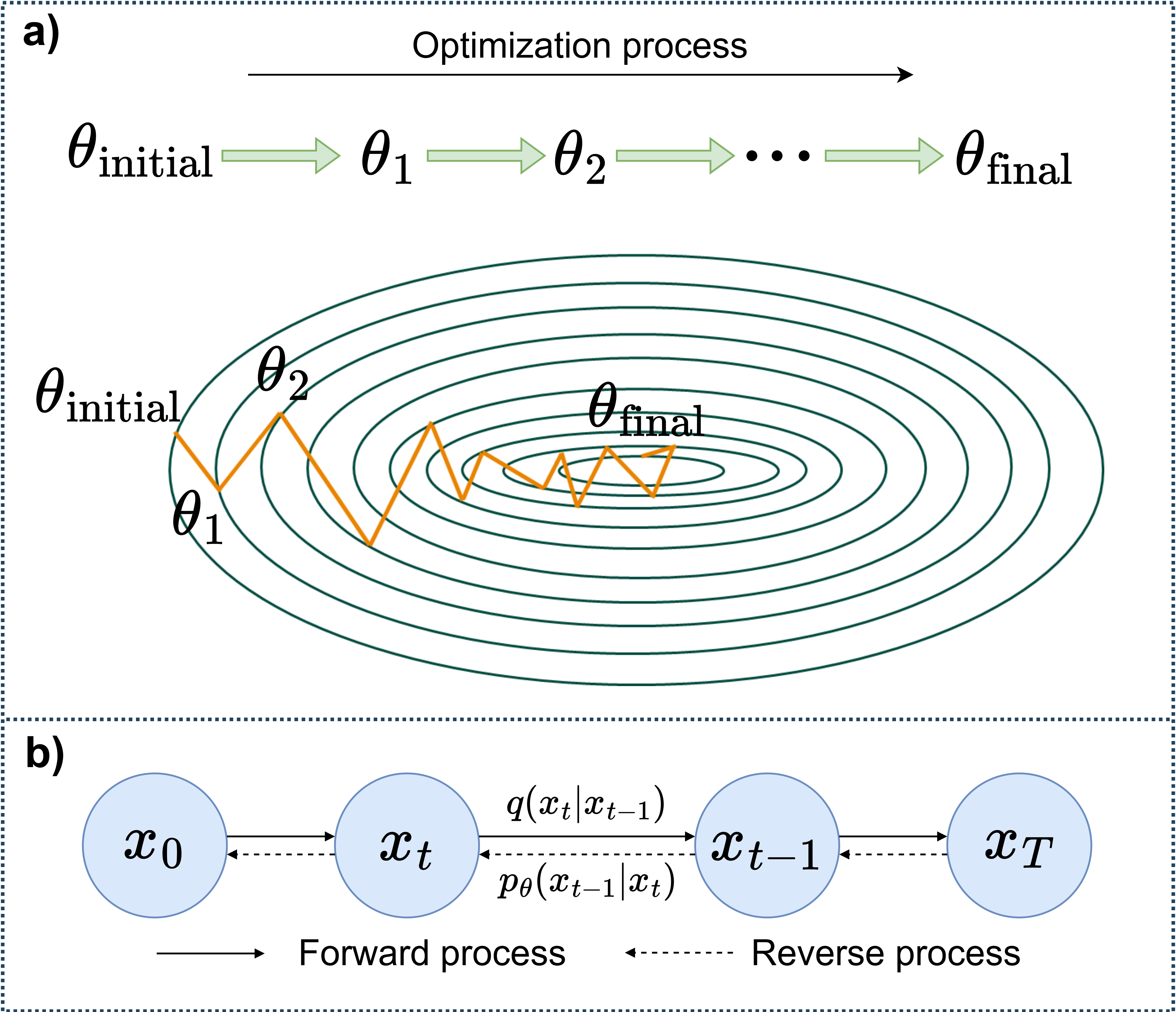}
\caption{(a)~Illustration of the PINN optimization process, starting from a random initialization \(\boldsymbol{\theta}_{\text{initial}}\) and converging to \(\boldsymbol{\theta}_{\text{final}}\). (b)~Overview of a generative diffusion model, which performs a forward process to add noise to \(x_0\) and a reverse process to denoise \(x_T\).}
\label{fig1}
\end{figure}

\subsection{PINNs for seismic wavefield representation}
PINNs aim to approximate the solution $\mathbf{u}(\mathbf{x}, t)$ of a governing wave equation at spatial-temporal location $(\mathbf{x}, t)$ by embedding physical constraints directly into its loss function \citep{raissi2019physics}. Let $\boldsymbol{\theta}$ denote the network parameters. Starting from an initial state $\boldsymbol{\theta}_{\mathrm{initial}}$, which is commonly drawn randomly, PINN training proceeds iteratively via gradient-based optimization until convergence, yielding a final set $\boldsymbol{\theta}_{\mathrm{final}}$ (see Figure \ref{fig1}a.). Formally, each update can be written as
\begin{equation}\label{eq1}
    \boldsymbol{\theta}_{k+1}
    = 
    \boldsymbol{\theta}_{k}
    - 
    \eta \,\nabla_{\boldsymbol{\theta}}\,\mathcal{L}(\boldsymbol{\theta}_{k}),
\end{equation}
where $\mathcal{L}$ comprises the PDE residual, boundary/initial conditions, and any regularization terms, while $\eta$ denotes the learning rate. Although this procedure ultimately yields a final parameter set $\boldsymbol{\theta}_{\mathrm{final}}$ for accurate wavefield modeling, relying on purely random initialization often leads to slow convergence, especially for complex velocity models. 

\subsection{Generative diffusion models}
GDMs provide an effective way to synthesize new samples, such as images, from a learned distribution and it does that by gradually removing noise from a sample drawn from a Gaussian distribution \citep{ho2020denoising}. They involve a forward process that transforms clean data $x_0$ into progressively noisier versions $x_1, x_2, \dots, \mathbf{x}_T$, often modeled as
\begin{equation}\label{eq2}
   q(x_t \mid x_{t-1}) \;=\; 
    \mathcal{N}\bigl(\sqrt{\alpha_t}\,x_{t-1},\;(1-\alpha_t)\mathbf{I}\bigr),
\end{equation}
where $\alpha_t$ controls how much noise is added at each time step $t$, and $q(x_t \mid x_{t-1})$ represents the conditional distribution of the sample $x_t$, at time step t, given $x_{t_1}$. The conditional distribution of the reverse process is given by
\begin{equation}\label{eq3}
    p_{\boldsymbol{\theta}}(x_{t-1} \mid x_t) \;=\; 
    \mathcal{N}\!\bigl(x_{t-1};\;\mu_{\boldsymbol{\phi}}(x_t, t),\;\Sigma_{\boldsymbol{\phi}}(x_t, t)\bigr),
\end{equation}
which iteratively recovers clean data from noise. Here, $\mu_{\phi}(x_{t}, t)$ and $\Sigma_{\phi}(x_{t}, t)$ are the mean and variance predicted by the trained diffusion model parameterized by $\boldsymbol{\phi}$. As depicted in Figure~\ref{fig1}b, starting from $x_T$ (random noise), a trained diffusion model applies repeated denoising steps to generate $x_0$. This approach has proved highly successful in tasks ranging from image synthesis to speech processing.

\subsection{The connection betwen PINN optimization and reverse diffusion process}
Although PINNs and GDMs are originally developed for different purposes, the training of PINNs and the reverse process of GDMs share an interesting conceptual resemblance. In PINN optimization (Figure~\ref{fig1}a), we begin with a random initialization, $\boldsymbol{\theta}_{\text{initial}}$, and progressively refine the parameters to minimize the loss $\mathcal{L}(\boldsymbol{\theta})$. By contrast, GDMs (Figure~\ref{fig1}b) start from random noise $x_T$ and iteratively remove noise until yielding a clean sample $x_0$. 

Both processes can be seen as an evolution from a highly disordered or random state toward a structured, physically (or visually) meaningful products. In PINNs, the structure emerges as the network parameters adapt to the governing equations and also the physical properties, like the velocity. In GDMs, structure appears when the noise is reversed according to the learned denoising distribution. If we consider the PINN's parameter initialization $\boldsymbol{\theta}_{\mathrm{initial}}$ (randomly drawn) as gaussian noise, then iteratively updating it into $\boldsymbol{\theta}_{\mathrm{final}}$ resembles a denoising trajectory. So, the PINN parameter update is a form of denoising that can be performed with a pretrained conditioned PINN that uses the velocity model to help obtain a good approximate $\boldsymbol{\theta}$. This insight suggests that if we learn a GDM over optimized PINN parameters, we could generate well-initialized parameters for new velocity models by sampling from the learned distribution.

\section{\textbf{Method}}\label{sec:method}
\begin{figure}
\centering
\includegraphics[width=0.98\textwidth]{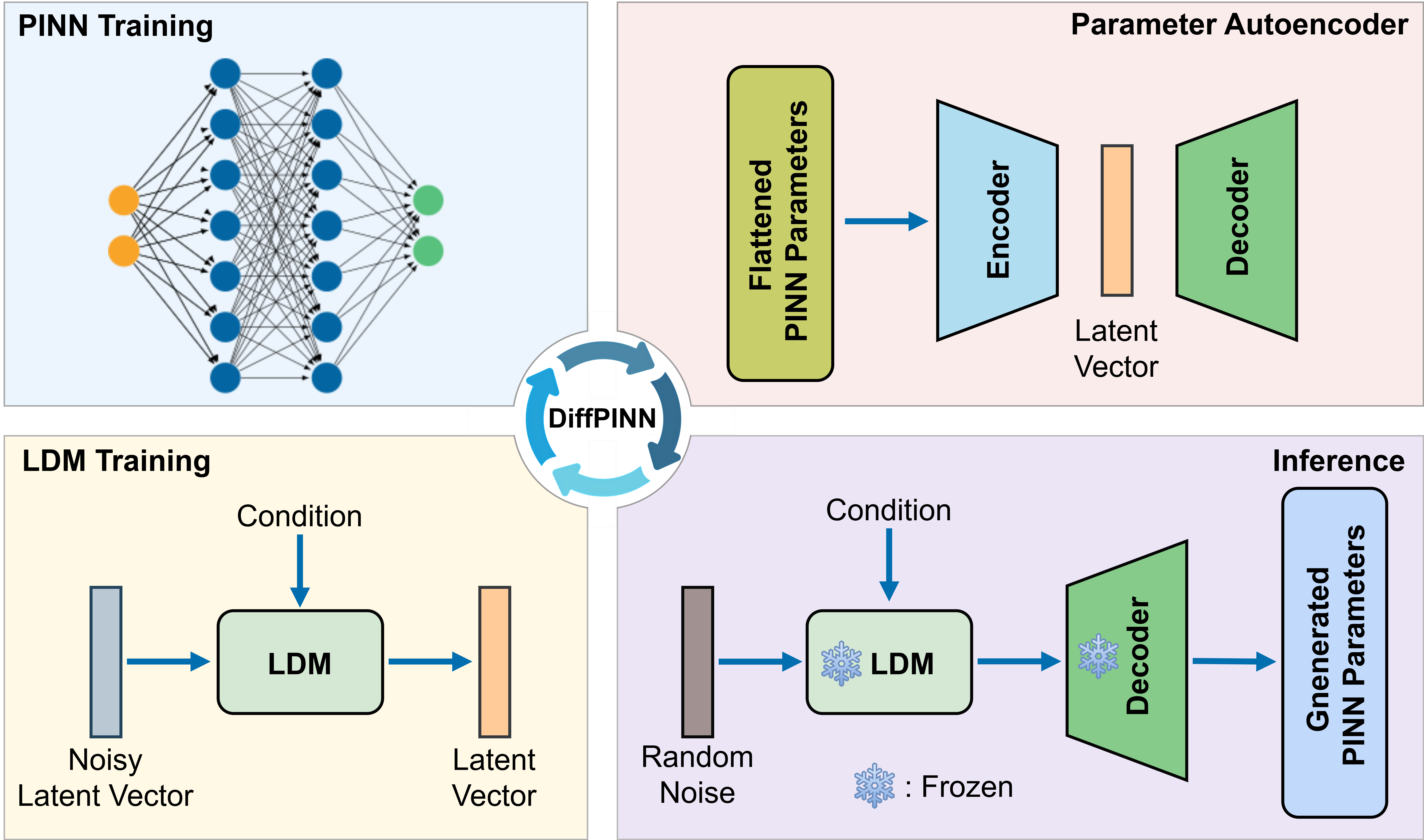}
\caption{An overview of our latent diffusion approach for generating PINN initialization parameters. 
    (a) PINN Training: We collect converged parameter vectors from PINNs trained under diverse velocity models. 
    (b) Autoencoder Training: We learn an encoder--decoder pair $(E,D)$ to compress these high-dimensional parameters into low-dimensional latent vectors. 
    (c) Latent Conditional Diffusion Training: We fit a diffusion model to the latent vectors, using velocity and source embeddings as conditioning inputs. 
    (d) Inference: For a new velocity model, we sample a latent vector via the diffusion reverse process (conditioned on the new model) and decode it to obtain a high-quality initial parameter set for PINN training.}
\label{fig2}
\end{figure}

In this section, we will detail how we borrow ideas from GDMs to generate improved initialization parameters for PINNs, a key motivation explored in this work. As illustrated in Figure~\ref{fig2}, the main idea is to collect multiple sets of trained PINN parameters, compress them into a low-dimensional latent space via an autoencoder, and then train a conditional diffusion model to learn the distribution of these latent representations. Once trained, this model can generate new parameter initializations tailored for previously unseen velocity models, significantly reducing the training cost of PINNs. We decompose our method into four stages: (1) PINN training, (2) Autoencoder training, (3) Conditional diffusion training, and (4) Inference.

\subsection{PINN Training for the scattered wavefield solutions}
Here, we focus on using PINN to represent frequency-domain scattered wavefields. Following the approach proposed by \cite{alkhalifah2021wavefield}, the scattered wavefield $\delta u(\mathbf{x},\mathbf{x}_s,\omega)$ satisfies the following perturbation equation:
\begin{equation}\label{eq4}
\omega^2\,m(\mathbf{x})\,\delta u(\mathbf{x},\mathbf{x}_s,\omega) 
\;+\; 
\nabla^2 \,\delta u(\mathbf{x},\mathbf{x}_s,\omega)
\;=\;
-\omega^2\,\delta m(\mathbf{x}) \, u_0(\mathbf{x},\mathbf{x}_s,\omega),
\end{equation}
where $\omega$ denotes the angular frequency, \(u_0(\mathbf{x},\mathbf{x}_s,\omega)\) is the background wavefield in a homogeneous medium, \(\mathbf{x}_s\) denotes the source location, \(m(\mathbf{x}) = 1/v^2\) represents the squared slowness of the medium, \(m_0(\mathbf{x}) = 1/v_0^2\) is the constant squared slowness in the background medium of velocity \(v_0\), and \(\delta m(\mathbf{x}) = m(\mathbf{x})-m_0(\mathbf{x})\). When the background medium is homogeneous and infinite, the reference wavefield \(u_0(\mathbf{x},\mathbf{x}_s,\omega)\) can be obtained analytically. For a 2D case, it takes the form \citep{Aki1980QuantitativeST}
\begin{equation}\label{eq5}
u_0(\mathbf{x},\mathbf{x}_s,\omega) 
\;=\; 
\frac{\mathrm{i}}{4}\,
H_0^{(2)}\!\Bigl(\tfrac{\omega}{v_0}
\,\bigl\lvert \mathbf{x}-\mathbf{x}_s \bigr\rvert\Bigr),
\end{equation}
where \(H_0^{(2)}\) denotes the zero-order Hankel function of the second kind, and \(\mathrm{i}\) is the imaginary unit. Equation~\eqref{eq5} thus provides a fast analytical solution for the background wavefield at any spatial position.

We train the network by enforcing the physics of equation \eqref{eq4} on collocation training samples \(\{\mathbf{x}^j, \mathbf{x}_s^j\}\) in the computational domain. More concretely, the physics-based loss is defined as
\begin{equation}\label{eq6}
\mathcal{L}_{\mathrm{phys}}(\boldsymbol{\theta})
\;=\;
\frac{1}{N}\sum\limits_{j=1}^{N} 
\Bigl\lVert 
\nabla^2\,\delta u(\mathbf{x}^j,\mathbf{x}_s^j,\omega;\boldsymbol{\theta})
\;+\;
\omega^2\,m(\mathbf{x}^j)\,\delta u(\mathbf{x}^j,\mathbf{x}_s^j,\omega;\boldsymbol{\theta})
\;+\;
\omega^2\,\delta m(\mathbf{x}^j)\,u_0(\mathbf{x}^j,\mathbf{x}_s^j,\omega)
\Bigr\rVert^2.
\end{equation}
Through gradient-based optimization (e.g., AdamW), the network parameters \(\boldsymbol{\theta}\) converge to a solution that satisfies the scattered wavefield equation. 

We repeat this scattered wavefield PINN training process for a diverse set of velocity models $\bigl\{v^{(i)}\bigr\}$, $i=1,\dots,M,$ to capture different velocity settings. Upon convergence, each trained PINN yields a final parameter vector \(\boldsymbol{\theta}^{(i)}\). Collecting these vectors, we form a dataset
\begin{equation}\label{eq7}
   \bigl\{\boldsymbol{\theta}^{(1)},\,\boldsymbol{\theta}^{(2)},\dots,\boldsymbol{\theta}^{(M)}\bigr\},
\end{equation}
which serves as the training to our subsequent autoencoder and diffusion models. However, directly applying a diffusion model to these high-dimensional parameter vectors remains computationally expensive. Hence, we first compress them into low-dimensional latent representations via an autoencoder, as described next.

\subsection{Autoencoder training}\label{subsec:autoencoder_training}
To address the high dimensionality of the parameter space, we employ an autoencoder, which comprises an encoder $E$ and a decoder $D$, to reduce each parameter vector from the training set into a much lower-dimensional latent representation. Specifically, for each converged parameter vector $\boldsymbol{\theta}^{(i)}$, the encoder produces:
\begin{equation}\label{eq8}
    \mathbf{z}^{(i)} \;=\; E\bigl(\boldsymbol{\theta}^{(i)}\bigr),
\end{equation}
while the decoder reconstructs it via
\begin{equation}\label{eq9}
    \hat{\boldsymbol{\theta}}^{(i)} \;=\; D\bigl(\mathbf{z}^{(i)}\bigr).
\end{equation}
We train $(E, D)$ by minimizing the mean squared error (MSE):
\begin{equation}\label{eq10}
    \mathcal{L}_{\mathrm{AE}} 
    \;=\;
    \sum_{i}
    \Bigl\lVert 
      \boldsymbol{\theta}^{(i)} 
      - 
      D\bigl(E(\boldsymbol{\theta}^{(i)})\bigr) 
    \Bigr\rVert^{2}.
\end{equation}

By selecting a suitable latent dimension, we preserve the key features necessary for accurate wavefield representation while discarding less relevant parameter variations. This dimensionality reduction not only eases computational requirements for the subsequent diffusion training but can also smooth out minor irregularities in the parameters, effectively providing a form of regularization. 

\subsection{Conditional diffusion training in latent space}\label{subsec:diffusion_training}
Once we have transformed the PINN parameters into a latent space, we train a conditional diffusion model to learn the distribution of these latent vectors. Following standard diffusion formulations, let $\mathbf{z}_0$ denote a latent vector (corresponding to a single $\boldsymbol{\theta}^{(i)}$) in the training set, and let $\mathbf{z}_t$ be its corrupted version at diffusion step $t$. During the forward process, noise is gradually added:
\begin{equation}\label{eq11}
    q(\mathbf{z}_t \mid \mathbf{z}_{t-1}) \;=\; 
    \mathcal{N}\bigl(\sqrt{\alpha_t}\,\mathbf{z}_{t-1},\;(1-\alpha_t)\mathbf{I}\bigr),
\end{equation}
where $\alpha_t$ controls the noise level at each step. The reverse process, parameterized by a network with parameters~$\boldsymbol{\phi}$, aims to invert this corruption by predicting the original latent vector $\mathbf{z}_0$ at each step (often referred to as ``$x_0$ prediction'' in the diffusion literature):
\begin{equation}\label{eq12}
    p_{\boldsymbol{\phi}}\bigl(\mathbf{z}_{t-1}\mid \mathbf{z}_{t}, \mathbf{c}\bigr)
    \;=\;
    \mathcal{N}\Bigl(
      \mathbf{z}_{t-1};\, 
      \mu_{\boldsymbol{\phi}}(\mathbf{z}_{t}, \mathbf{c}, t),\, 
      \Sigma_{\boldsymbol{\phi}}(\mathbf{z}_{t}, \mathbf{c}, t)
    \Bigr).
\end{equation}
Here, $\mathbf{c}$ represents explicit conditioning information that directs the generation toward the appropriate latent representations. 

In our context, $\mathbf{c}$ includes two main elements: (1)~velocity sample coordinates and their associated velocity values, and (2)~source coordinates used in the PINN training process. Let $\mathbf{x}$ and $\mathbf{x}_s$ denote, respectively, the sampled spatial points in the velocity model and the source coordinates. We embed these coordinates through a small network $f_{\mathrm{coord}}(\cdot)$, yielding
\begin{equation}\label{eq13}
   \mathbf{e}^{(\mathbf{x})}
   =
   f_{\mathrm{coord}}(\mathbf{x}),
   \quad
   \mathbf{e}^{(\mathbf{x}_s)}
   =
   f_{\mathrm{coord}}(\mathbf{x}_s).
\end{equation}
If the sampled velocity at point $\mathbf{x}$ is denoted $v(\mathbf{x})$, we similarly embed this value using a small network $f_{\mathrm{vel}}(\cdot)$:
\begin{equation}\label{eq14}
   \mathbf{e}^{(v)}
   =
   f_{\mathrm{vel}}\!\bigl(v(\mathbf{x})\bigr).
\end{equation}
We then concatenate these embeddings into a single conditional vector:
\begin{equation}\label{eq15}
   \mathbf{c}
   \;=\;
   \mathrm{Concat}\Bigl(
     \mathbf{e}^{(\mathbf{x})},\; \mathbf{e}^{(\mathbf{x}_s)},\;
     \mathbf{e}^{(v)}
   \Bigr).
\end{equation}
This $\mathbf{c}$ is fed into each residual block of the diffusion network, allowing the denoising trajectory to take into account both the velocity field and the source configuration. Consequently, the final denoised latent vector aligns with parameters suitable for modeling seismic waves using PINN under the specified conditions. 

In contrast to predicting noise at each step, we follow an $x_0$ (or $\mathbf{z}_0$) prediction strategy, which several prior studies have shown to improve both training convergence and generation quality \citep{bansal2024cold}. Therefore, our diffusion model is trained to output an estimate $\hat{\mathbf{z}}_0$ of the original latent vector $\mathbf{z}_0$ at each diffusion step. We train by minimizing the single latent-space MSE:
\begin{equation}\label{eq16}
    \mathcal{L}_{\mathrm{diff}} 
    \;=\;
    \mathbb{E}_{\mathbf{z}_0,\,t}\Bigl\lVert
      \hat{\mathbf{z}}_0(\mathbf{z}_t, \mathbf{c}, t)
      \;-\;
      \mathbf{z}_0
    \Bigr\rVert^2.
\end{equation}
This loss alone encourages the diffusion model to recover accurate latent codes that, when decoded, yield high-quality PINN parameter initializations.   

\subsection{Inference: Generating new PINN parameters with physics guidance}\label{subsec:inference}
At inference stage, we aim to obtain PINN initialization parameters for a new, previously unseen velocity model $v_{\text{new}}$. We begin by constructing its conditional embedding $\mathbf{c}_{\text{new}}$ via the same embedding functions $f_{\mathrm{coord}}$ and $f_{\mathrm{vel}}$:
\begin{equation}\label{eq17}
   \mathbf{c}_{\text{new}}
   =
   \mathrm{Concat}\Bigl(
     f_{\mathrm{coord}}(\mathbf{x}),\; 
     f_{\mathrm{vel}}\bigl(v_{\text{new}}(\mathbf{x})\bigr),\;
     f_{\mathrm{coord}}(\mathbf{x}_s)
   \Bigr).
\end{equation}

Then, we sample a noise vector $\mathbf{z}_T$ from a standard Gaussian distribution $\mathcal{N}(\mathbf{0},\mathbf{I})$ in latent space and iteratively apply the learned reverse diffusion from $t=T$ down to $t=1$. At each time step $t$, we first sample an intermediate latent
\begin{equation}\label{eq18}
    \mathbf{z}_{t-1} 
    \;\sim\; 
    p_{\boldsymbol{\phi}}\bigl(\mathbf{z}_{t-1}\mid \mathbf{z}_{t},\,\mathbf{c}_{\mathrm{new}}\bigr).
\end{equation}

Next, to inject physics guidance, we decode $\mathbf{z}_{t-1}$ through the pretrained decoder $D$ to obtain a full parameter vector
\begin{equation}\label{eq19}
    \hat{\boldsymbol{\theta}} 
    = D(\mathbf{z}_{t-1}).
\end{equation}
We then evaluate the physics-based PINN loss $\mathcal{L}_{\mathrm{phys}}(\hat{\boldsymbol{\theta}})$ on the same collocation samples used in training (i.e., Equation ~\eqref{eq6}). By backpropagating this loss through $D$, we compute the gradient with respect to the intermediate latent $\mathbf{z}_{t-1}$:
\begin{equation}\label{eq20}
    \nabla_{\mathbf{z}_{t-1}}\;\mathcal{L}_{\mathrm{phys}}\bigl(D(\mathbf{z}_{t-1})\bigr)\,.
\end{equation}

After this, we can correct the latent sample via a small gradient step:
\begin{equation}\label{eq21}
    \mathbf{z}_{t-1}
    \;\leftarrow\;
    \mathbf{z}_{t-1}
    \;-\;
    \gamma\,
    \nabla_{\mathbf{z}_{t-1}}\;\mathcal{L}_{\mathrm{phys}}\bigl(D(\mathbf{z}_{t-1})\bigr),
\end{equation}
where $\gamma$ is a step size (like the learning rate). The corrected $\mathbf{z}_{t-1}$ is then used as the starting point for the next diffusion step. 

Once completing all $T$ denoising steps, we can obtain $\mathbf{z}_{0}^{\text{new}}$, which should ideally lie close to the manifold of latent vectors corresponding to accurate PINN parameters for $v_{\text{new}}$. Finally, we decode $\mathbf{z}_{0}^{\text{new}}$ using the trained decoder $D$ to obtain the complete (flattened) parameter vector:
\begin{equation}\label{eq22}
    \boldsymbol{\theta}^{\text{new}} 
    \;=\; 
    D\bigl(\mathbf{z}_{0}^{\text{new}}\bigr).
\end{equation}
This parameter vector is done by restoring the shape of each linear layer parameter of PINN and, thus, can serve as the initial point for training a PINN under the new velocity model. 
These parameters, already imbued with both learned diffusion priors and physics-based correction, serve as a powerful initialization for subsequent PINN training, leading to faster convergence and higher solution fidelity than starting from a random initialization.

\begin{figure}[htbp]
\centering
\includegraphics[width=1\textwidth]{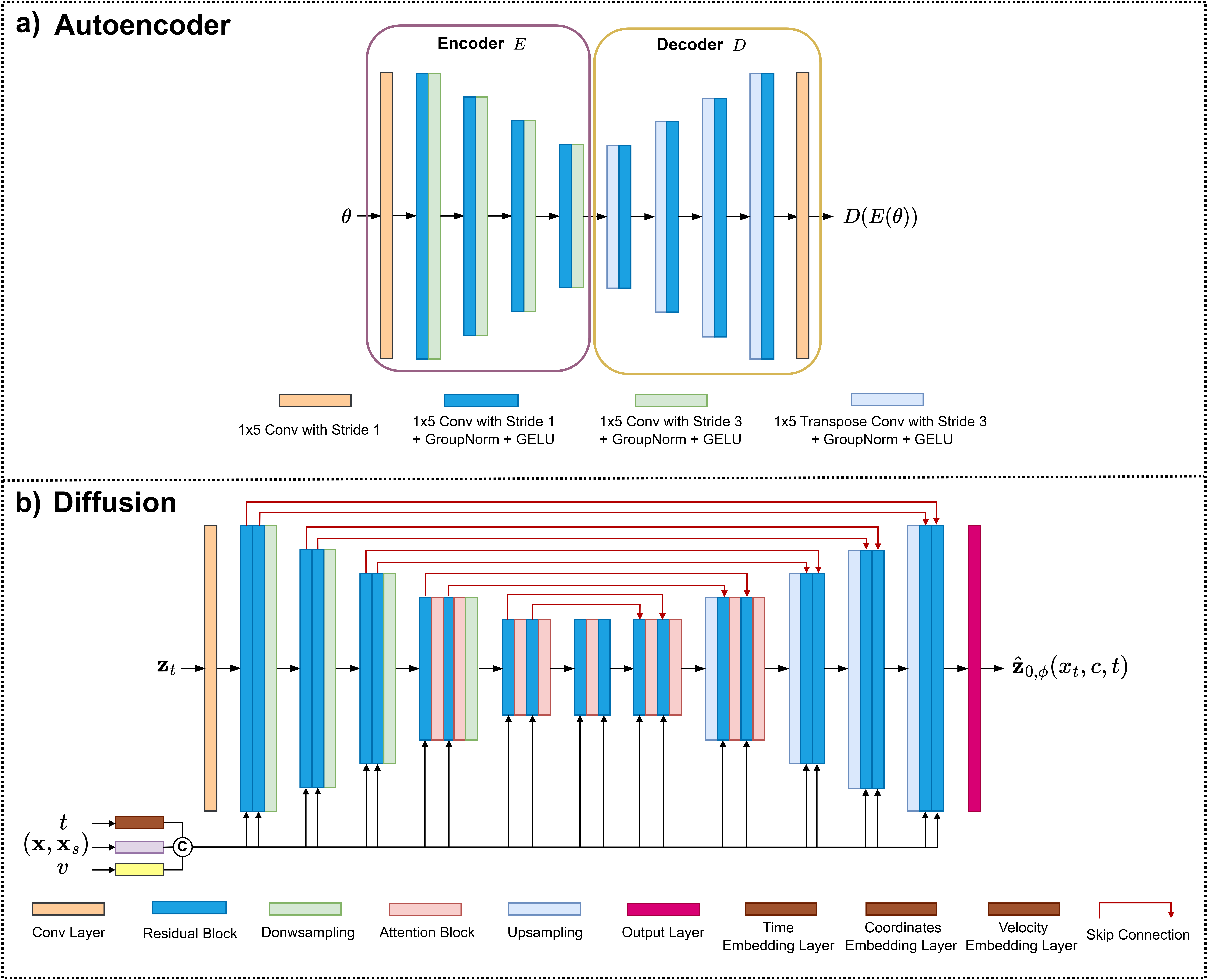}
\caption{An illustration of network architectures. (a) Autoencoder compresses the flatted PINN parameter vector \(\boldsymbol{\theta}\) into a \(128\times1590\) latent vector $\mathbf{z}$. (b) Conditional diffusion U-Net, where the input to the network is the noisy latent vector $\mathbf{z}_t$, and the conditions, $\mathbf{c}$, and the output is the denoised latent vector, $\hat{\mathbf{z}}_0$, at time step, $t$.}
\label{network}
\end{figure}

\subsection{Network architecture}\label{subsec:network_architectures}
We employ three main neural networks (NNs) in our approach: the PINN for wavefield modeling, the autoencoder for dimensionality reduction, and the diffusion model for latent vector generation. 

Our PINN is a multi-layer perceptron (MLP) comprising six hidden layers with neurons \(\{256, 256, 128, 128, 64, 64\}\) from the shallow to deeper layers, and uses a $\sin$ activation function. This design yields a flattened parameter vector of size \(1 \times 128770\).  

Our autoencoder use a 1D convolutional NN (CNN), which is illustrated in Figure~\ref{network}a. We first applies a \(1\times5\) convolution (stride 1) to the flattened parameter vector \(\boldsymbol{\theta}\in\mathbb{R}^{1\times128770}\), expanding from 1 to 64 channels The encoder then proceeds through four stages. In each stage, we first apply a \(1\times5\) convolution with stride 1, group normalization (GroupNorm), and Gaussian error linear units (GELU) activation function, and then a second \(1\times5\) convolution with stride 3 for downsampling, again followed by GroupNorm and GELU. After these four stages, the channel counts at the output of each downsampling are exactly \(\{64,128,128,128\}\), yielding the latent tensor \(\mathbf{z}\in\mathbb{R}^{128\times1590}\). The decoder mirrors this process in four upsampling stages. At each stage, we first apply a \(1\times5\) transpose convolution (stride 3) to upsample spatially, followed by GroupNorm and GELU, then a \(1\times5\) convolution (stride 1) with GroupNorm and GELU to refine features. The channel counts after each transpose-convolution are \(\{128,512,512,64\}\), and the subsequent stride-1 convolutions preserve these same channel counts. Finally, a last \(1\times5\) convolution (stride 1) reduces from 64 channels back to 1 channel, reconstructing \(\hat{\boldsymbol{\theta}}\). 

Our diffusion model adopts a 1D U-Net design that progressively denoises the latent code \(\mathbf{z}_t\in\mathbb{R}^{128\times1590}\) under conditioning \(\mathbf{c}\) (see Figure~\ref{network}b). We first apply a \(1\times3\) convolution (stride 1) that maps the latent input to 128 feature maps. The encoder then consists of five sequential stages with channel widths \(\{128,256,512,1024,1024\}\). In stages 1$\sim$3, each stage applies two residual blocks, followed by a downsampling block that halves the spatial dimension and doubles the channel count. In stage 4, we augment each residual block with a self-attention module (applied immediately after the block), then perform the downsampling step, yielding 1024 channels. Stage 5 omits the downsampling but retains the two residual+attention units. At the network’s deepest point, a bottleneck applies a residual block, a self-attention block, and a second residual block, all at 1024 channels. The decoder symmetrically reverses this process: starting from 1024 channels, each of its five stages first upsamples (via a transpose convolution that doubles the spatial resolution and halves the channels) and then applies two residual blocks with embedded self-attention (in the stage corresponding to encoder 4). Skip connections link each encoder stage’s output to the corresponding decoder stage’s input. Finally, a \(1\times3\) convolution maps the 128-channel feature maps back to the latent estimate \(\hat{\mathbf{z}}_0\). Time, coordinates, and velocity embeddings are added into every residual block to guide the denoising according to the conditioning vector \(\mathbf{c}\). 

All three networks, including PINN, autoencoder, and diffusion model, are described in more detail in our open-source repository: \url{https://github.com/DeepWave-KAUST/DiffPINN}.
\section{\textbf{Numerical examples}}
In the following, we present numerical experiments to validate the effectiveness and efficiency of our proposed DiffPINN framework. The experiments are organized as follows. First, we describe our training configurations, including details about datasets, training procedures, and hyperparameters. Next, we evaluate the performance of DiffPINN on velocity models sampled within the training distribution (in-distribution tests). Subsequently, we assess its generalization capability on velocity models outside the training distribution (out-of-distribution tests). Finally, we compare DiffPINN's performance between in-distribution and out-of-distribution scenarios for more specific generalization understanding.

\subsection{Training configuration}
For training, we extract 2600 distinct velocity models from the CurveVel-A class of the OpenFWI dataset \citep{deng2022openfwi}. Each model originally has a resolution of $70 \times 70$, where we resize each to $101 \times 101$ and apply mild smoothing to each velocity model. The grid spacing in both $x$ and $z$ directions is set to 25 m. For each velocity model, we randomly sample 20000 points for training, which include the spatial coordinates, the corresponding velocity, the source location, and a background velocity model. 

Training 2600 separate PINNs from scratch, one per velocity model, is computationally expensive. To mitigate this, we build on our previous Meta-PINN approach \citep{cheng2025meta}, which provides a robust initialization via meta-learning. Specifically, we select 500 of the 2600 velocity models for meta-training and optimize a meta-network for 40000 iterations. In each iteration, we randomly sample 10 training tasks (velocity models) from the 500 meta-training models, splitting them equally into support and query sets. The inner-loop learning rate is fixed at $2\times 10^{-3}$, while the outer-loop learning rate starts at $1.5\times 10^{-3}$ and is decayed by 0.8 every 5000 iterations. We then use this meta-learned initialization to train all 2600 PINNs for 15000 iterations each, ensuring that they converge to high-fidelity solutions. All PINN training is conducted on ten NVIDIA V100 GPUs (32 GB each) and took approximately 65 hours in total. The meta-learning initialization also helped guide the network to structurally similar local minima solutions for the various velocity models.

Once all 2600 PINNs are fully trained, we flatten their parameters into vectors. These vectors form the training data for a 1D autoencoder, which is trained for 1000 epochs with a batch size of 64. The initial learning rate is $1\times 10^{-3}$, and it is decayed by 0.8 at 100, 250, 500, and 750 epochs. After completing the autoencoder training, we use its encoder to obtain latent representations for each of the 2600 PINN parameter sets. We then train a 1D diffusion model on these latent vectors with a fixed learning rate of $5\times 10^{-5}$, a batch size of 20, and an exponential moving average (EMA) rate of 0.999. The diffusion model is trained for 400000 iterations on an NVIDIA A100 GPU (80 GB) and took approximately 23.5 hours. All network training employs an AdamW optimizer \citep{loshchilov2017decoupled}. 

In our subsequent in-distribution and out-of-distribution tests, we employ the denoising diffusion implicit models (DDIM) sampler \citep{song2020denoising} with only 10 reverse diffusion steps to substantially reduce the time required for PINN parameter generation. To ensure a fair and general evaluation, we use a fixed PINN training schedule for all experiments: an initial learning rate of $1.5\times10^{-3}$, decayed by a factor of 0.6 at epochs 2000, 4000, 6000, and 8000. As baselines, we also optimize PINNs starting from the meta-learned initialization (denoted by Meta-PINN) and from a standard random initialization (hereafter vanilla PINN). All velocity models, including in-distribution and out-of-distribution, maintain a uniform grid spacing of 25 m, and we randomly select 20000 points from each model for PINN training. 

\subsection{Test on in-distribution velocity models}
To evaluate performance and, also, ensure generalization, we select five new velocity models, not one, from the CurveVel-A class (distinct from those used in training). For each of these five test models, we use our trained diffusion model to generate the corresponding latent representation and, then, use the decoder of the trained autoencoder to obtain the initialization parameters of each PINN and, thus, to perform further optimization. 

\begin{figure}[htbp]
\centering
\includegraphics[width=0.6\textwidth]{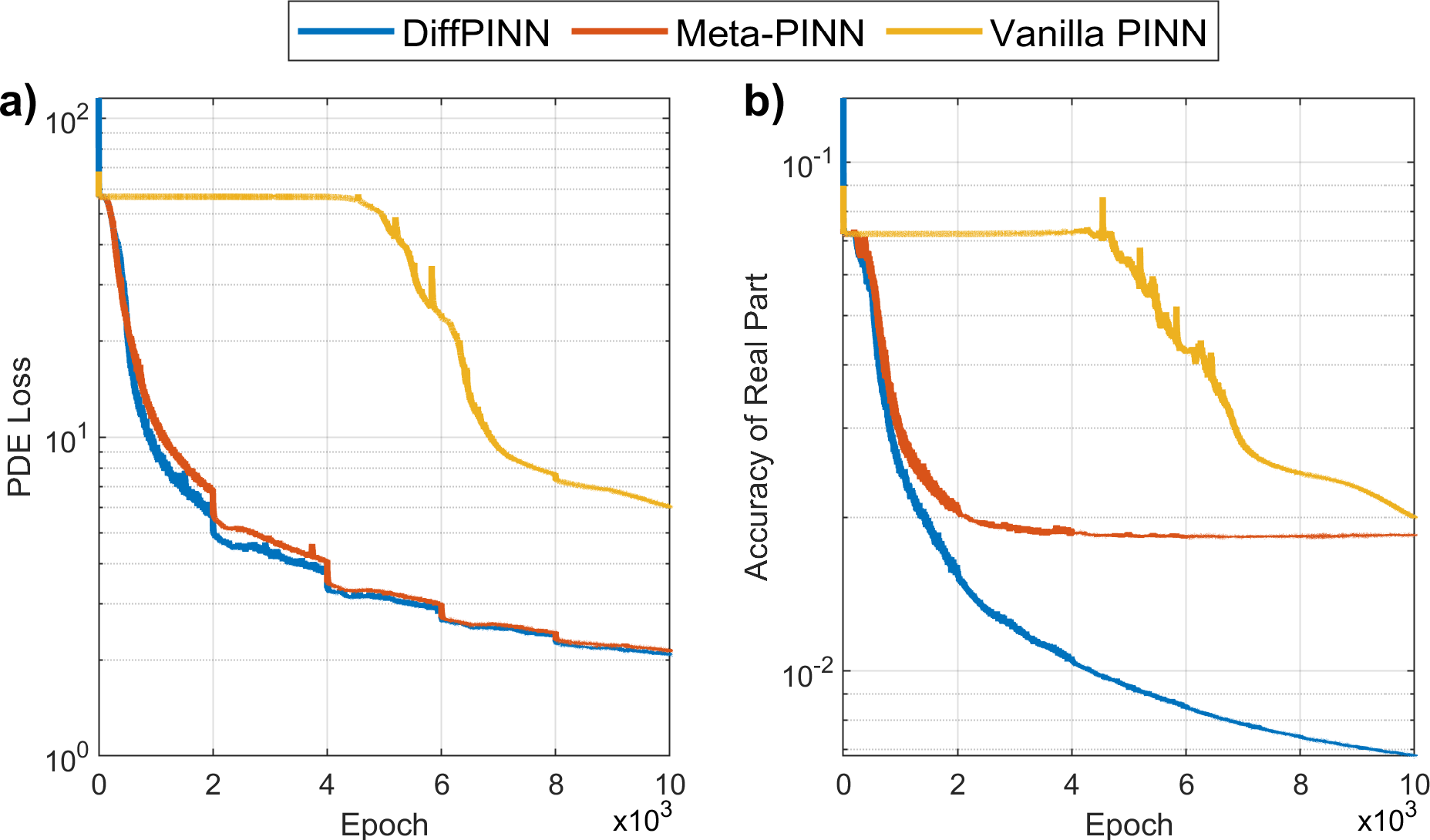}
\caption{Performance comparison among DiffPINN (blue), Meta-PINN (orange), and vanilla PINN (yellow), averaged over the five in-distribution velocity models. (a) The averaged physical loss curves. (b) The averaged accuracy curves of the real part of the scattered wavefield solutions relative to numerical reference solutions. }
\label{fig3}
\end{figure}

Figure~\ref{fig3} shows the averaged physical loss and accuracy curves over the five tested in-distribution velocity models. In panel (a), we can observe that DiffPINN achieves slightly lower PDE loss compared to Meta-PINN, indicating a modest improvement in convergence speed, while both methods significantly outperform the vanilla PINN. Vanilla PINN exhibits a long plateau in the initial training phase, reflecting its difficulty in finding a reliable descent direction early on, which is a key factor behind its slower convergence. Panel (b) illustrates the accuracy of the real part of the scattered wavefield relative to the numerical reference solution (in terms of MSE). Here, and in subsequent test, we omit displaying accuracy curves for the imaginary part of the scattered wavefield, as they exhibit trends very similar to the real part and, thus, would provide redundant information. DiffPINN consistently demonstrates significantly higher accuracy compared to Meta-PINN, highlighting a clear advantage. Meanwhile, the vanilla PINN exhibits considerably lower accuracy, emphasizing the inherent challenges of PINNs and underscoring the importance of an effective initialization strategy.  

\begin{figure}[htbp]
\centering
\includegraphics[width=0.98\textwidth]{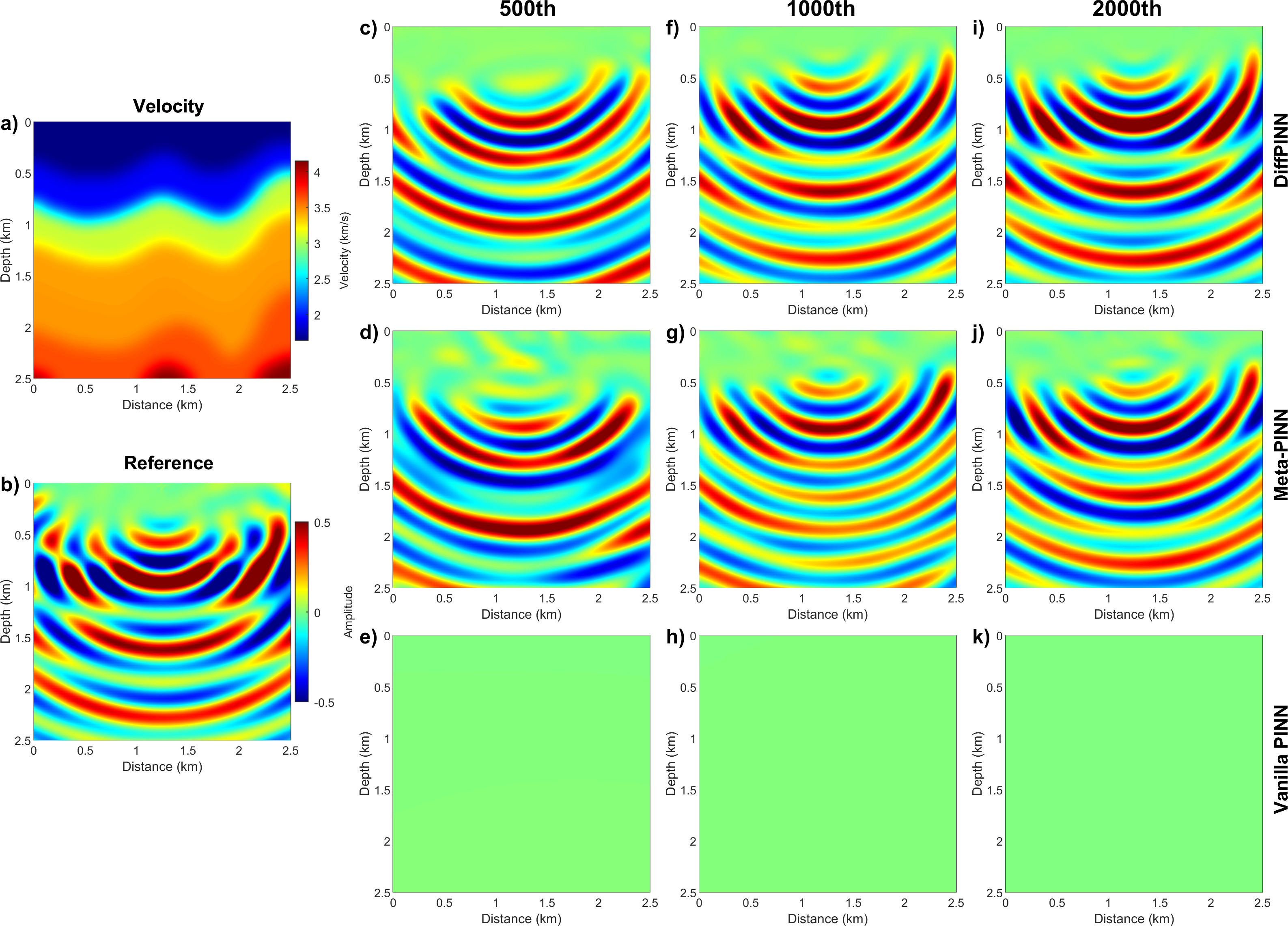}
\caption{Comparison of the real part of the scattered wavefield solutions at 5 Hz for an in-distribution test velocity model. (a) Velocity model. (b) Numerical reference solution. Subsequent rows, from top to bottom, represent wavefields predicted by DiffPINN, Meta-PINN, and vanilla PINN, respectively. Columns correspond to different training epochs, where the specific epoch numbers are indicated in the top.}
\label{fig4}
\end{figure}

Figure~\ref{fig4} shows a detailed comparison of the real part of the scattered wavefield solutions obtained by DiffPINN and the two benchmark methods for one representative velocity model among the five test cases. Panel~(a) displays the selected velocity model, while panel~(b) shows the corresponding numerical reference solution computed by FD method. The remaining panels present the predicted wavefields by DiffPINN (top row), Meta-PINN (middle row), and vanilla PINN (bottom row) at epochs 500, 1000, and 2000 (from left to right). We can see that DiffPINN quickly captures the overall wavefield structure even within the first 500 epochs and refines the details as training proceeds, matching the reference wavefield closely by 2000 epoch. Meta-PINN also provides competitive wavefield representations, but it noticeably lags behind DiffPINN in capturing finer details, which explains its significantly lower accuracy seen earlier in the averaged accuracy curves. In contrast, the vanilla PINN fails to provide a reasonable wavefield solution even after 2000 epochs. These results confirm that DiffPINN offers a significant advantage in both convergence speed and final wavefield accuracy.

\subsection{Test on out-of-distribution velocity models}
To evaluate the robustness and generalization capability of our DiffPINN method to out-of-distribution velocity models, we select five additional velocity models significantly different from those used during training. These models include a layered velocity structure extracted from the Marmousi model (with an original grid size of $91\times91$), and four models selected from four distinct classes of the OpenFWI dataset: FlatFault-A, FlatFault-B, CurveFault-A, and CurveFault-B. The original resolution of these models is $70\times70$, which we resize to $101\times101$ for consistency. 

\begin{figure}[htbp]
\centering
\includegraphics[width=0.6\textwidth]{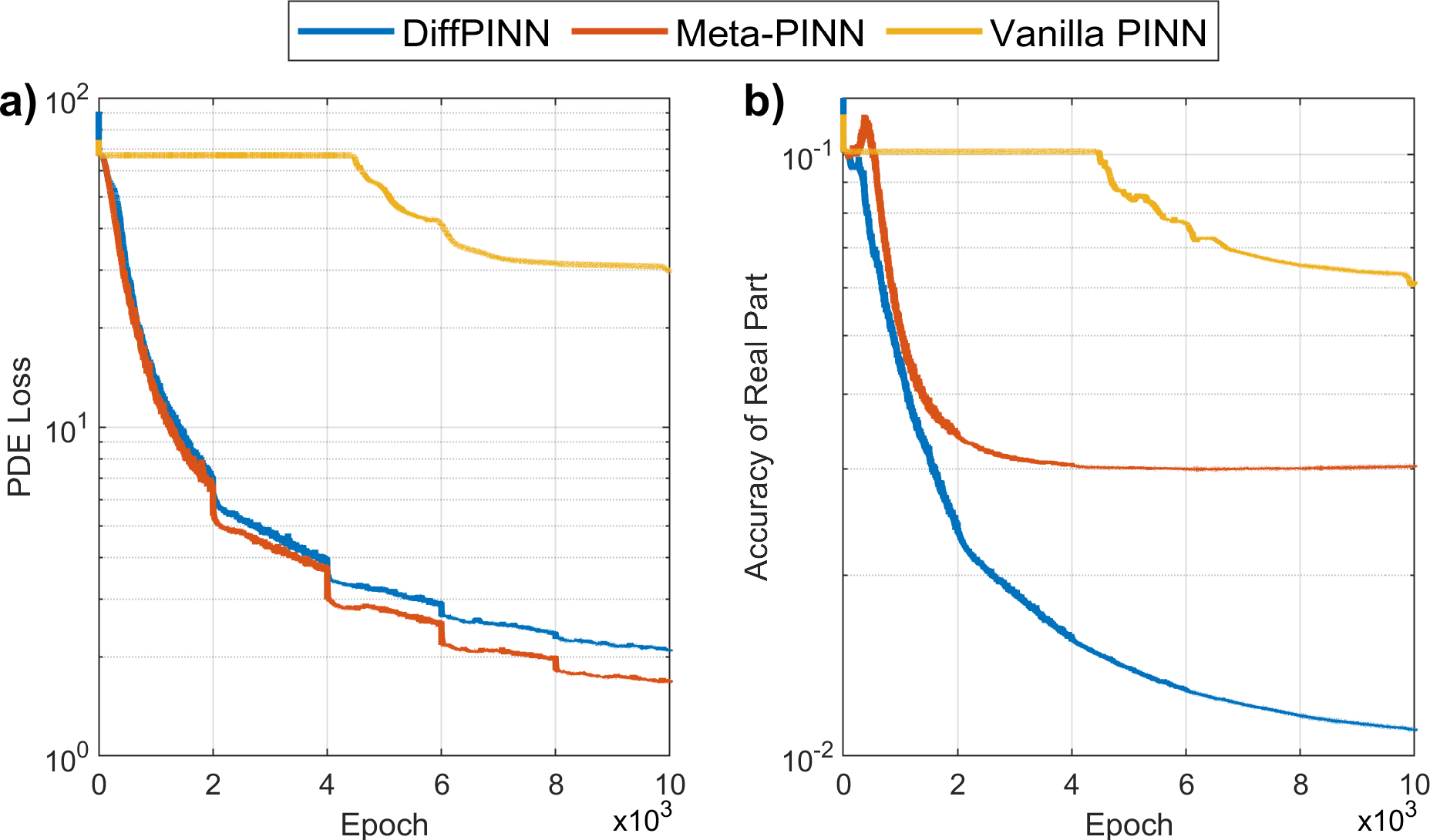}
\caption{Performance comparison among DiffPINN (blue), Meta-PINN (orange), and vanilla PINN (yellow), averaged over the five out-of-distribution velocity models. (a) The averaged physical loss curves. (b) The averaged accuracy curves of the real part of the scattered wavefield solutions relative to numerical reference solutions. }
\label{fig5}
\end{figure}

We first show the averaged physical loss and accuracy curves (real part of the scattered wavefield) across the five selected out-of-distribution models in Figure~\ref{fig5}. From the physical loss curves, we can observe that Meta-PINN initially exhibits faster convergence than DiffPINN. However, accuracy curves clearly reveal that DiffPINN consistently achieves significantly higher accuracy compared to Meta-PINN, similar to observations in the in-distribution tests. This apparent inconsistency between the physical loss and accuracy for Meta-PINN can be attributed to its convergence toward trivial or poor local minima solutions, thereby failing to capture meaningful wavefield details despite relatively lower PDE losses. In contrast, vanilla PINN shows notably slower convergence in both physical loss and accuracy curves, significantly underperforming compared to DiffPINN and Meta-PINN. 

\begin{figure}[htbp]
\centering
\includegraphics[width=0.98\textwidth]{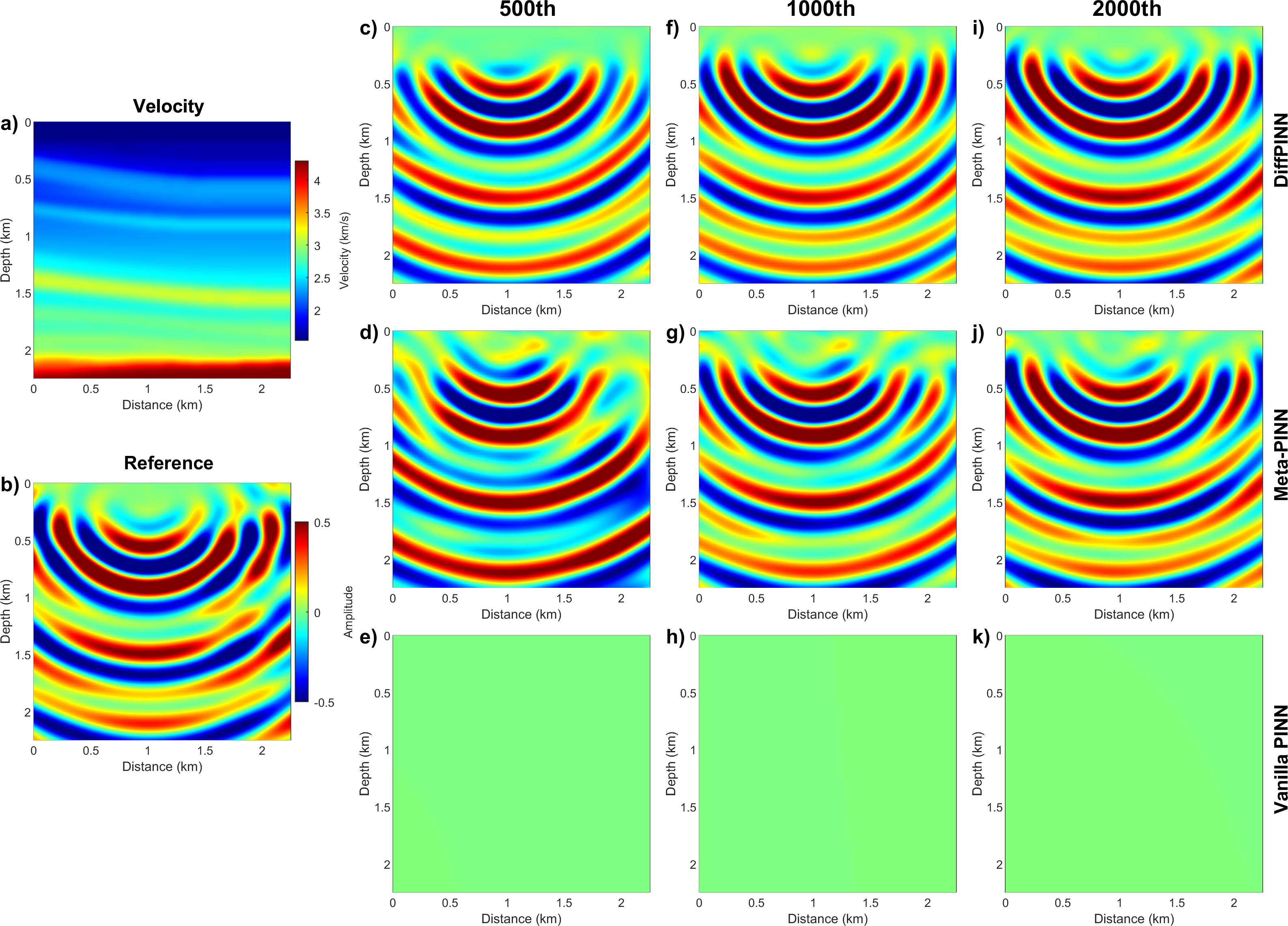}
\caption{Similar with Figure \ref{fig4}, but for layered velocity model extracted from the Marmousi model.}
\label{fig6}
\end{figure}

To further illustrate these observations, we select two representative velocity models and examine their predicted wavefields in detail. Figure~\ref{fig6} shows wavefield comparisons for the layered model extracted from Marmousi. Panels in Figure~\ref{fig6} follow the same layout as Figure~\ref{fig4}. DiffPINN accurately captures the overall wavefield structure as early as epoch 500, progressively refining finer details through subsequent epochs. By epoch 2000, DiffPINN provides wavefield predictions closely matching the numerical reference solution. In contrast, Meta-PINN captures the general structure at epoch 500 but clearly struggles to represent shallow wavefield features accurately. Even after 2000 epochs, Meta-PINN predictions exhibit notable differences from the reference in fine-scale details. Vanilla PINN completely fails to provide a meaningful wavefield representation, even after training up to epoch 2000. 

\begin{figure}[htbp]
\centering
\includegraphics[width=0.98\textwidth]{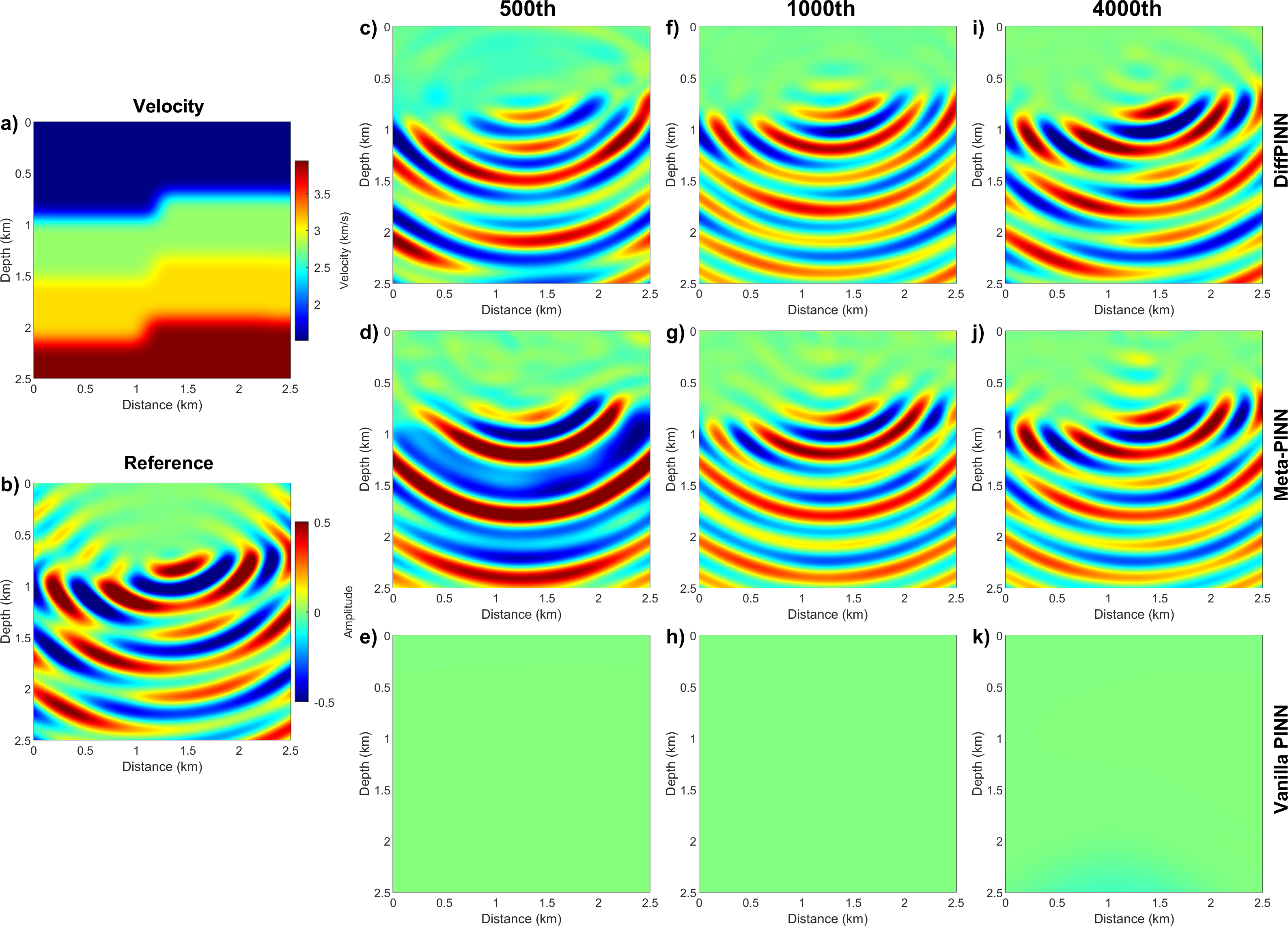}
\caption{Similar with Figure \ref{fig4}, but for FlatFault-A velocity model.}
\label{fig7}
\end{figure}

Figure~\ref{fig7} further presents wavefield comparisons for the out-of-distribution FlatFault-A velocity model. Panel arrangements again follow the structure established in Figure~\ref{fig4}, except that the rightmost column corresponds to epoch 4000. DiffPINN again demonstrates superior capability in capturing both the wavefield structure and amplitude more accurately than Meta-PINN from epoch 500 onwards. As training progresses, DiffPINN increasingly matches fine-scale details such as wavefield discontinuities induced by faults. Conversely, Meta-PINN struggles with these detailed wavefield features, providing only approximate structural representations even at epoch 4000. Vanilla PINN, once again, fails to yield any meaningful wavefield solutions, highlighting its inadequacy in the absence of effective initialization.

\begin{figure}[htbp]
\centering
\includegraphics[width=0.6\textwidth]{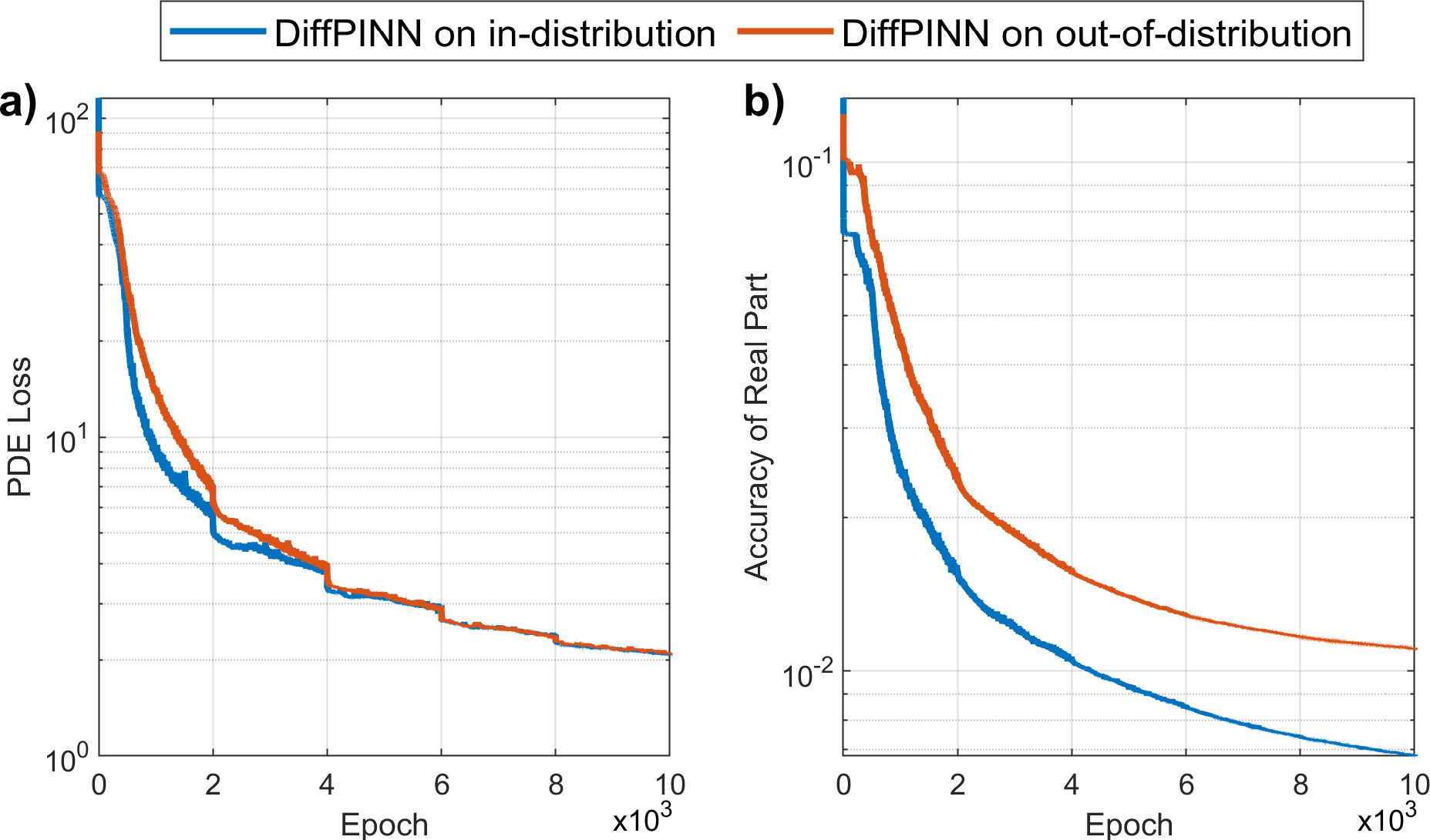}
\caption{Comparison of DiffPINN performance on in-distribution (blue) versus out-of-distribution (orange) velocity models. (a) The averaged physical loss curves. (b) The averaged accuracy curves of the real part of scattered wavefield relative to numerical reference solutions. }
\label{fig8}
\end{figure}

\subsection{In- vs. Out-of-distribution performance
analysis}
Although the results in previous the subsection demonstrate that DiffPINN generalizes well to out-of-distribution velocity models compared to Meta-PINN, it is important to quantify any performance degradation resulting from velocity distribution shifts. Figure~\ref{fig8} compares the averaged PDE loss and accuracy curves of real-part scattered wavefield of DiffPINN on the five in-distribution models versus the five out-of-distribution models. 

From panel~(a), we can observe that DiffPINN on out-of-distribution models incurs a small increase in PDE loss relative to in-distribution cases, indicating a slight slowdown in convergence. Similarly, panel~(b) shows a noticeable drop in accuracy for the real part of the scattered wavefield when moving out of distribution. This performance slip is expected: the diffusion model is trained to capture the latent parameter distribution of PINNs under the training velocity models, and a shift in velocity distribution can lead to a mismatch between the learned latent prior and the optimal initialization for new models. 

These results highlight the need to broaden the range of velocity models used during training. By incorporating a wider and more diverse set of velocity distributions, the diffusion model can learn a more comprehensive latent parameter prior, thereby enhancing DiffPINN’s generalization and robustness across even more varied subsurface scenarios.

\section{\textbf{Discussion}}
In this section, we further analyze and interpret the results of our numerical experiments, focusing on four key aspects: (1) what insights can be gained from the generated PINN weights; (2) the benefits of physics-guided parameter generation; (3) the impact of different diffusion sampling step sizes on parameter performance; and
(4) an outlook on the broader implications of this work.

\subsection{Insights from the generated weights} 
\begin{figure}[htbp]
\centering
\includegraphics[width=1\textwidth]{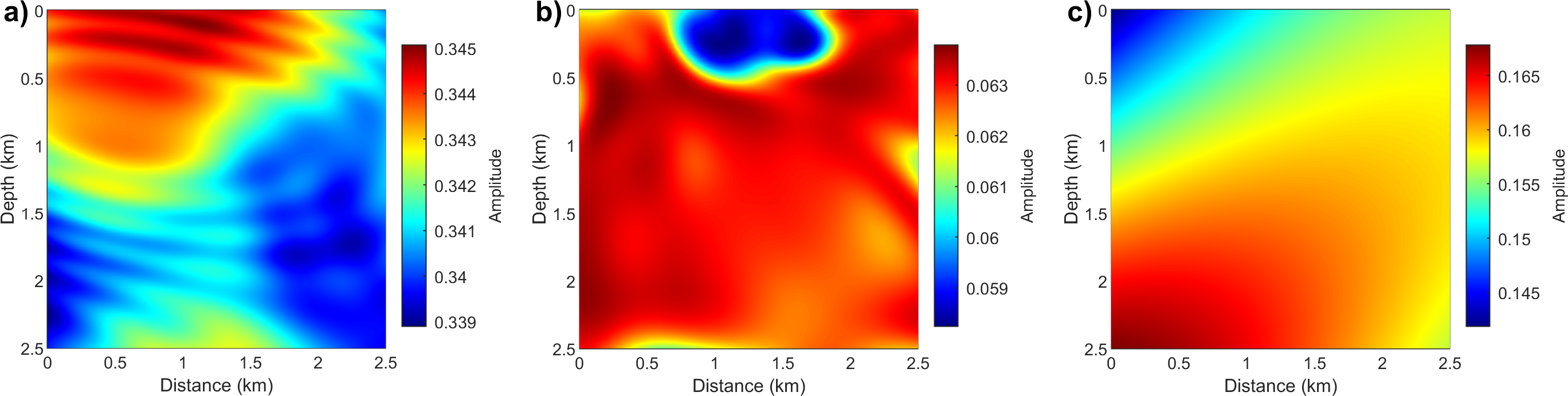}
\caption{Initial real part of the scattered wavefield predicted directly from the different weights (epoch 0): (a) DiffPINN’s generated weights, (b) Meta-PINN’s weights, and panel, and (c) random initialization (vanilla PINN). Here, we use the test in Figure~\ref{fig4} as an example.}
\label{fig9}
\end{figure}

Figure~\ref{fig9} compares the initial real part of the scattered wavefield predicted directly from the generated weights (epoch 0) for the test in Figure~\ref{fig4}. Panel (a) shows the wavefield from DiffPINN’s generated weights, panel (b) from Meta-PINN initialization, and panel (c) from random initialization (vanilla PINN). We can see that:
\begin{itemize}
    \item Vanilla PINN yields an overly smooth wavefield lacking physical detail.
    \item Meta-PINN captures some variation, particularly near the source, but remains coarse compared to the reference.
    \item DiffPINN produces richer fine-scale structure and significantly larger amplitudes, indicating that its generated weights encode meaningful physical patterns even before any training.
\end{itemize}

\begin{figure}[htbp]
\centering
\includegraphics[width=1\textwidth]{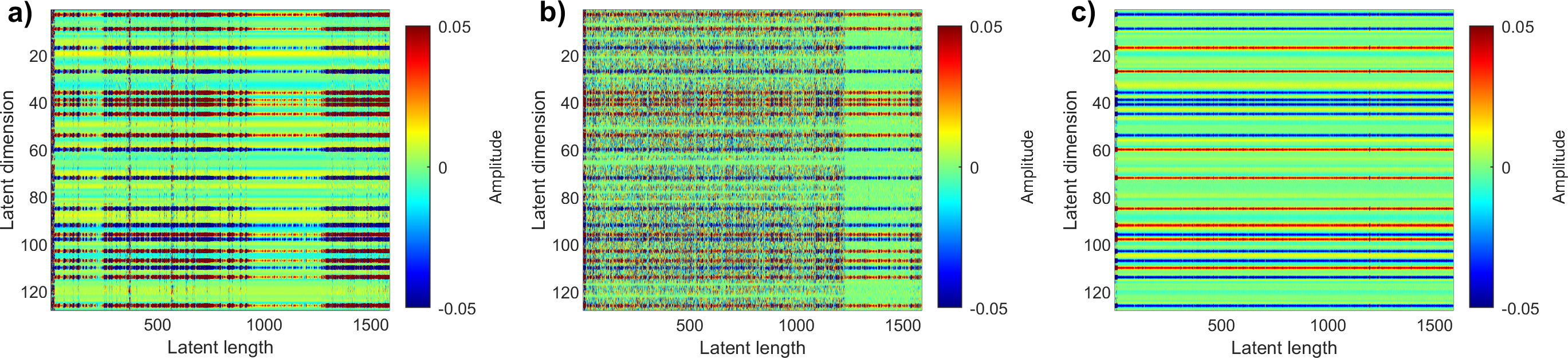}
\caption{Latent representations \(\mathbf{z}_0\) of the weights for each initialization method. (a), (b), and (c) correspond to DiffPINN, Meta-PINN, and vanilla PINN, respectively. Each latent map is of size \(128 \times 1590\), with the vertical axis indexing latent dimensions and the horizontal axis indexing latent vector length.}
\label{fig10}
\end{figure}

To understand these differences in more depth, Figure~\ref{fig10} visualizes the corresponding latent representations \(\mathbf{z}_0\) for each initialization. Panels (a), (b), and (c) show the 128 × 1590 latent maps for DiffPINN, Meta-PINN, and vanilla PINN, respectively. The vertical axis indexes the 128 latent dimensions and the horizontal axis spans the 1590-length vector. Key observations include:
\begin{itemize}
    \item Vanilla PINN latent map appear remarkably clean in the latent space, showing horizontal bands along certain latent dimensions. This is likely because the bottleneck of the trained autoencoder forces it to retain only those features that repeatedly occur in the training data and aid reconstruction. In contrast, random initialization, which typically samples from a zero-mean Gaussian (or uniform) distribution, lacks such reconstructible patterns, so it’s treated as noise and completely filtered out, leaving behind a smooth, nearly constant latent representation.
    \item Meta-PINN latent map exhibits strong random noise on the left, overlaid on faint banded structures. This suggests meta-learning retains unstable components for cross-task adaptability, while the emerging horizontal bands on the right reflect optimization directions beneficial across tasks, likely shaped by the training velocity distribution.
    \item DiffPINN latent map removes the noise and displays pronounced high-frequency variations along horizontal bands rather than being stationary. These concentrated patterns could correspond to a small number of key parameter directions that encode fine wavefield details, explaining why DiffPINN’s initial weights yield more detailed information (Figure~\ref{fig9}a) and enable faster convergence during subsequent training.
\end{itemize}

\subsection{The benefits of physics-guidance parameter generation}
To quantify the advantage of our physics-guided sampling strategy during diffusion inference, we compare two variants of our method: (1) physics-guided DiffPINN, which is detailed in Method Section~\ref{subsec:inference}, and (2) unguided DiffPINN, an otherwise identical conditional diffusion model that omits the physics-based gradient correction, relying solely on the learned latent prior and conditioning inputs. For both variants, we generate initial weights for ten test velocity models (five in-distribution and five out-of-distribution) and train the PINNs under the same hyperparameters. 

\begin{figure}[htbp]
\centering
\includegraphics[width=0.6\textwidth]{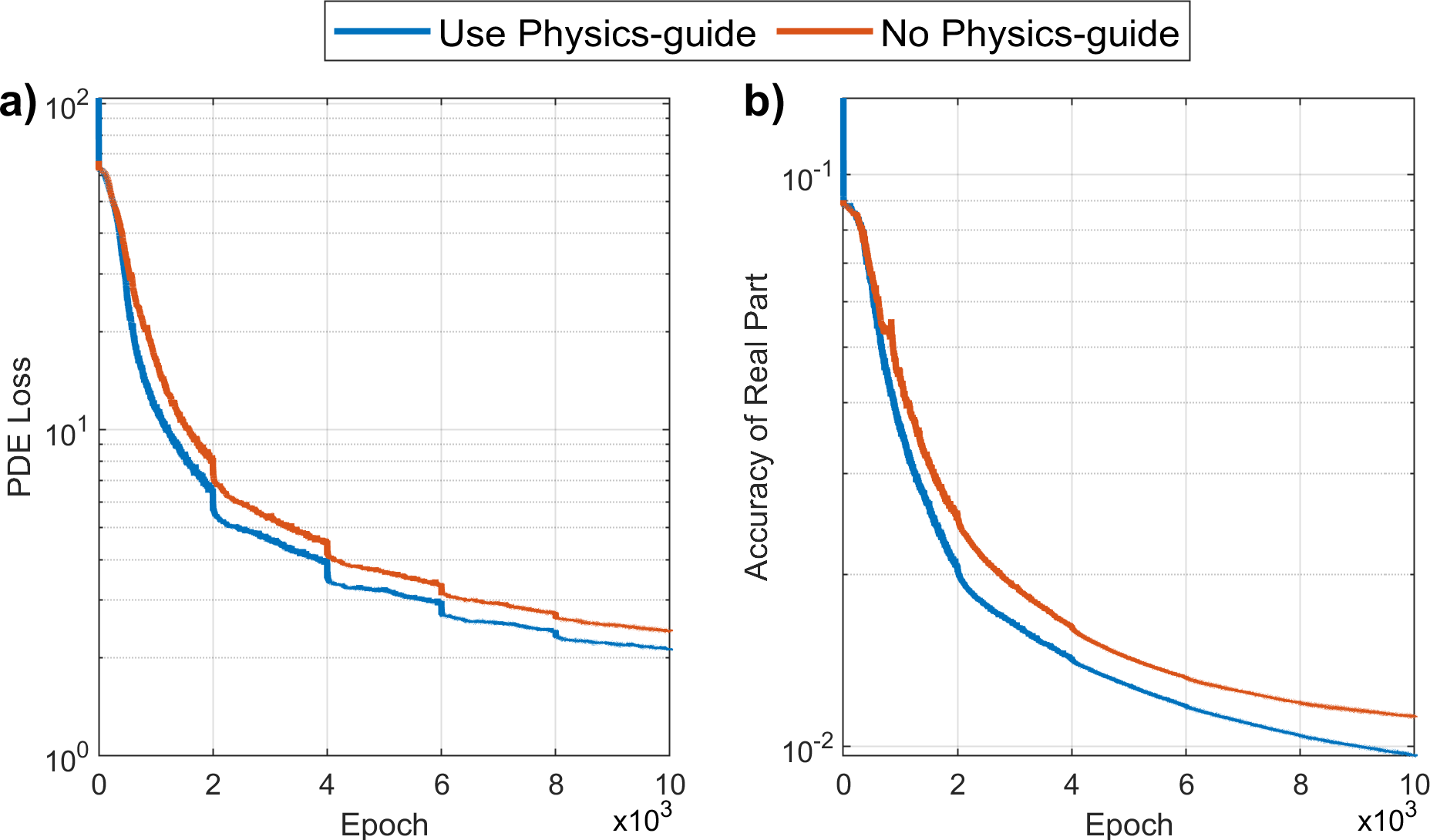}
\caption{Performance comparison between physics-guided and unguided generated initializations, averaged over five in-distribution and five out-of-distribution velocity models. (a) The averaged physical loss curves. (b) The averaged accuracy curves for the real part of the scattered wavefield.}
\label{fig11}
\end{figure}

Figure~\ref{fig11} presents the averaged physical loss (panel a) and accuracy curves of the real-part scattered wavefield (panel b), computed over all ten test models. We can see that PINNs initialized with physics-guided weights reach lower PDE loss significantly earlier than those with unguided weights. Across both in-distribution and out-of-distribution cases, physics-guided initialization yields consistently higher accuracy of the scattered wavefield throughout training. These results demonstrate that embedding physical context directly into the diffusion inference process not only accelerates optimization but also improves solution fidelity, underscoring the critical role of physics-guided parameter generation.

\subsection{The effect of the sampling step size}
\begin{figure}[htbp]
\centering
\includegraphics[width=0.6\textwidth]{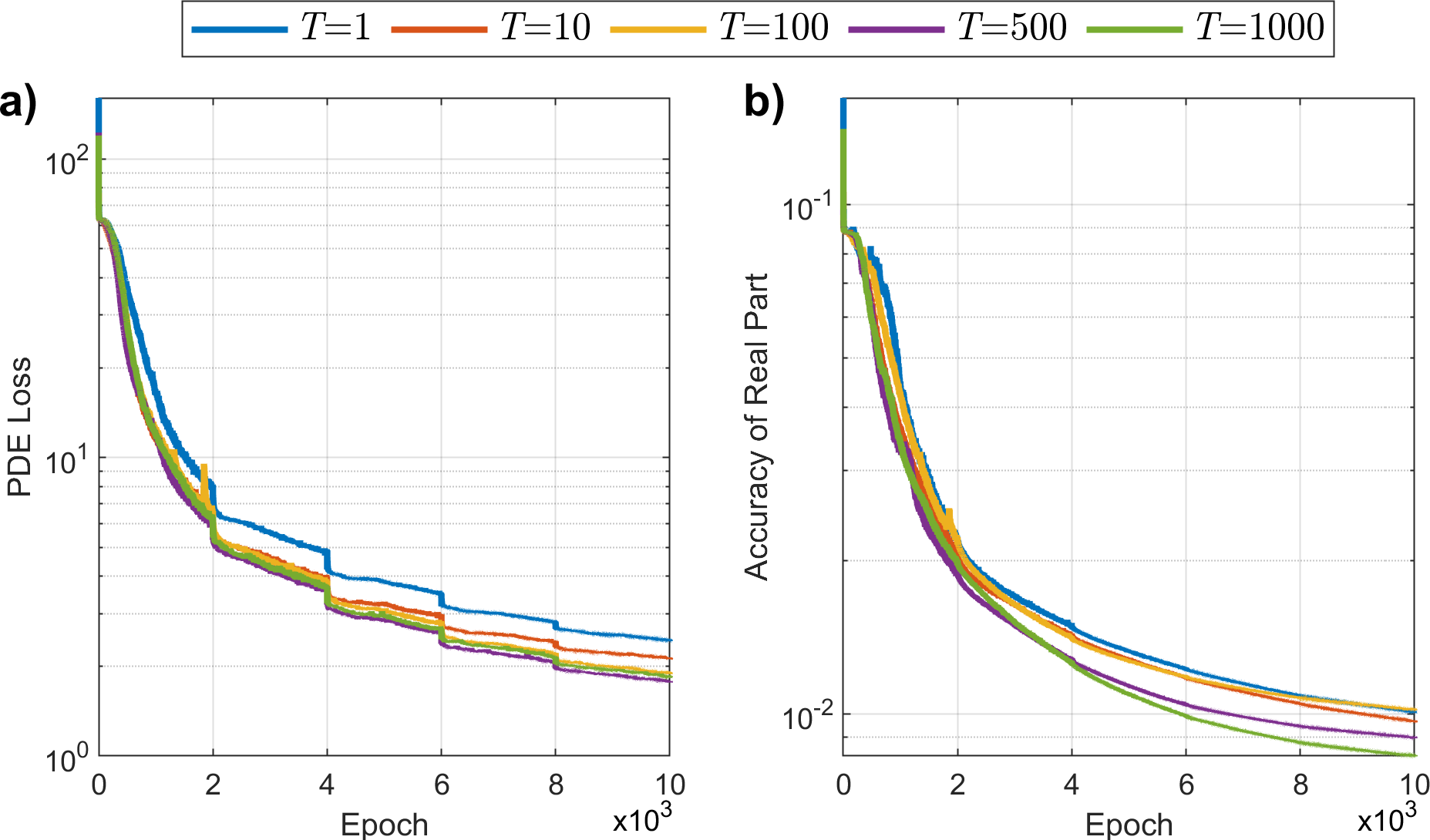}
\caption{Impact of reverse-diffusion sampling step size on PINN performance, averaged over five in-distribution and five out-of-distribution velocity models. (a) The averaged physical loss curves and (b) the averaged accuracy curves for sampling step sizes of 1, 10, 100, 500, and 1000.}
\label{fig12}
\end{figure}

We, here, investigate how the number of reverse-diffusion sampling steps affects the quality of generated PINN parameters. We generate weights for each of the 10 test velocity models (five in-distribution and five out-of-distribution) using sampling step sizes of 1, 10, 100, 500, and 1000. Each set of the generated weights is then used to initialize a PINN and train under the same hyperparameters described in the numerical experiments. We compute the averaged PDE loss and accuracy curves of the real-part scattered wavefield over all 10 models and plot them in Figure~\ref{fig12}. 

From Figure~\ref{fig12}(a), we can see that using only 1 sampling step yields the slowest convergence and highest final PDE loss, significantly worse than all other choices. Sampling step sizes of 100, 500, and 1000 produce nearly identical loss curves, each outperforming the 10-step configuration. In Figure~\ref{fig12}(b), the accuracy curves reveal a general trend: longer sampling produces higher solution accuracy, with 1000 steps achieving the best results. However, we sample time scales approximately linearly with the number of steps. For example, using 1000 steps requires nearly 100$\times$ the time of 10 steps. To balance computational cost and performance, we therefore adopt 10 sampling steps as our default configuration.

\subsection{Outlook and broader implications}
While we have demonstrated DiffPINN using a simple six-layer multilayer perceptron, our latent diffusion framework is agnostic to the underlying PINN architecture. Future work can leverage more advanced models, such as GaborPINN \citep{alkhalifah2024physics} or our previously proposed Lowrank-PINN \citep{cheng2025multi}, to generate initialization parameters that further accelerate convergence and improve accuracy. By providing high-quality priors for these richer architectures, diffusion-based initialization could unlock even greater efficiency gains in complex wavefield simulations. 

Beyond seismic wavefield modeling, the underlying idea has much broader significance and opens up multiple avenues of application:
\begin{itemize}
    \item \textbf{Other PINN-based PDE solvers.} Any PINN, whether solving fluid-dynamics equations, heat conduction, or electromagnetic problems, can benefit from diffusion-generated priors. As long as one can gather a set of trained PINN parameter vectors for varied scenarios, an analogous autoencoder+diffusion pipeline may be used to build a generative prior over network parameters. This perspective paves the way for accelerated PDE solution strategies across diverse PDE domains.
    \item \textbf{Implicit full-waveform inversion (IFWI).} IFWI methods that represent subsurface parameters via neural networks also suffer from lengthy iterative optimization. Our approach can be adapted to generate initial neuron representations for IFWI by learning the latent distribution of well-converged inversion networks. Sampling from this diffusion-trained prior would steer IFWI toward physically plausible parameter estimates from the start, substantially reducing iteration counts and overall runtime.
    \item \textbf{Neural-network (NN) based seismic processing.} NN-based seismic processing, such as denoising, interpolation, resolution enhancement, and velocity model building, often rely on fine-tuned deep networks tailored to specific datasets. By capturing the distribution of optimized network weights for these tasks, diffusion models can provide strong initializations when applying processing networks to new surveys or acquisition geometries. This warm start can shorten fine-tuning time, enabling rapid adaptation and deployment of advanced NN-based processing algorithms in real-time field operations.
\end{itemize}

\section{\textbf{Conclusions}}
We presented a novel latent diffusion-based approach to efficiently initialize physics-informed neural networks (PINNs) for seismic wavefield modeling. Our method leverages a two-step generative process: (1) training multiple PINNs on a diverse set of velocity models and compressing their final parameters into a low-dimensional latent space via an autoencoder, and (2) training a conditional diffusion model to store the distribution of these latent vectors conditioned by the velocity model, which allows for rapid sampling PINN parameters for new velocity models. Experimental results demonstrated that our framework converges substantially faster than both a meta-learned initialization and a standard random initialization, while achieving higher final accuracy in representing frequency-domain scattered wavefields on both in-distribution and out-of-distribution velocity models.

\section{\textbf{Acknowledgment}}
This publication is based on work supported by the King Abdullah University of Science and Technology (KAUST). The authors thank the DeepWave sponsors for their support. This work utilized the resources of the Supercomputing Laboratory at King Abdullah University of Science and Technology (KAUST) in Thuwal, Saudi Arabia.
\vspace{0.5cm}
\section{\textbf{Code Availability}}
The data and accompanying codes that support the findings of this study are available at: 
\url{https://github.com/DeepWave-KAUST/DiffPINN}. (During the review process, the repository is private. Once the manuscript is accepted, we will make it public.)

\bibliographystyle{unsrtnat}
\bibliography{references}

\end{document}